# Determination of Photonuclear Reaction Cross-Sections on stable p-shell Nuclei by Using Deep Neural Networks

Serkan Akkoyun[1,3,*], Hüseyin Kaya[1], Abdulkadir Şeker[2,3], Saliha Yeşilyurt[2,3]

[1]*Department of Physics, Faculty of Science, Sivas Cumhuriyet University, Sivas, Turkey*
[2]*Department of Computer Engineering, Faculty of Engineering, Sivas Cumhuriyet University, Sivas, Turkey*
[3]*Artificial Intelligence Systems and Data Science Application and Research Center, Sivas Cumhuriyet University, Sivas, 58140, Turkey*

**Abstract**

The photonuclear reactions which is induced by high-energetic photon are one of the important type of reactions in the nuclear structure studies. In this reaction, a target material is bombarded by photons with the energies in the range of gamma-ray energy scale and the photons can statistically be absorbed by a nucleus in the target material. In order to get rid of the excess energies of the excited target nuclei, it can first emit protons, neutrons, alphas and light particles according to the separation energy thresholds. After this emitting process, generally an unstable nucleus can be formed. By the investigation of this products forming after photonuclear reactions, nuclear structure information can be obtained. In the present work, (γ, n) photonuclear reaction cross-sections on stable p-shell nuclei have been estimated by using neural network method. The main purpose of this study is to find neural network structures that give the best estimations on the cross-sections and to compare them with each other and available literature data. According to the results, the method is convenient for this task.

## 1. INTRODUCTION

In the nuclear structure studies, reactions induced by photons are one of the important tools. In these types of nuclear reactions, the target nuclei are bombarded by high-energetic photons and the photons can statistically be absorbed by a nucleus in the target material. Because of a nuclear process can be observed in the reaction, these are called as photonuclear reaction (Strauch, 1953). The excited nucleus emits proton, neutron, alpha and light particles first to get rid of excess energy. In the case of neutron emission after absorbing photons, the reaction is called as photo-neutron ($\gamma$, n) reaction. The character of the photo-neutron reaction is purely electromagnetic. Therefore, it can be used for understanding nucleon-nucleon interaction, collective motion of the nuclear matter and nuclear state excitation mechanisms. In about 15-30 MeV energy region, photonuclear reaction cross-sections are large and stable nuclei may be transmuted to short-lived or stable ones by using these reactions. The experimental studies on these reactions have begun in 1934 (Chadwick & Goldhaber, 1934) but there is still lack of existing data. Therefore, systematic studies of photonuclear reactions on different nuclei are needed (Serkan Akkoyun, Bayram, Dulger, vd., 2016).

The cross-section values for photo-neutron reactions for different isotopes and different energies are determined either experimentally or theoretically (Ishkhanov & Orlin, 2009; Utsuno vd., 2015). One of the most used theoretical codes for this purpose is TALYS computer code. TENDL database (Koning vd., 2019) is based on this code and other sources such as ENDF. The code is a system for the analysis and prediction of nuclear reactions. The basic objective behind its construction is the simulation of nuclear reactions that involve neutrons, photons, protons, deuterons, tritons, $^3$He- and alpha-particles, in the 1 keV-200 MeV energy range and for target nuclides of mass 12 and heavier. To achieve this, it is implemented a suite of nuclear reaction models into a single code system.

The one of the easiest ways to produce the radioactive isotopes is photo-neutron ($\gamma$, n) reaction. $^8$Be, $^9$B, $^{11}$C, $^{13}$Ne and $^{15}$O can be generated by using photo-neutron reactions performed on $^9$Be, $^{10}$B, $^{12}$C, $^{14}$Ne and $^{16}$O stable isotopes. Therefore, the information about the cross-sections on p-shell nuclei according to different energy values for these reactions is very important. In the literature, there is no complete data for all photon energies on the isotopes. In the present study, neural network methods have been employed for the prediction of ($\gamma$, n) reaction cross sections in different energies from reaction threshold energy values to 200 MeV on stable or long-lived p-shell isotopes. The available cross-section data are taken from TENDL-2019 library [6]. The methods generate the own outputs as close as the desired values. One of the advantages of the method is it does not need any relationship between input and output data variables. Another advantage of the method is that in case of missing data, it can complete missing data thanks to its learning ability. Therefore, one can confidently estimate the cross-section values for the given target and energy values which are not available in the literature. Recently, neural networks have been used in many fields in nuclear physics. Among them the studies performed by our group are developing nuclear mass systematic (Tuncay Bayram vd., 2014), obtaining fission barrier heights (Serkan Akkoyun, 2020), obtaining nuclear charge radii (S. Akkoyun vd., 2013), estimation of beta decay energies (Serkan Akkoyun vd., 2014), approximation to the cross sections of Z boson (Serkan Akkoyun & Kara, 2013; Kara vd., 2014), determination of gamma-ray angular distributions (Yildiz vd., 2018), adjustment of relativistic mean field model parameters (T. Bayram vd., 2018), neutron-gamma discriminations (S. Akkoyun, 2013; Yildiz & Akkoyun, 2013) and estimations of radiation yields for electrons in absorbers (Serkan Akkoyun, Bayram, & Yildiz, 2016).

## 2. MATERIAL and METHODS

NN (neural network) methods are very powerful mathematical tools for almost all problems which are based on the brain functionality and nervous system (Haykin, 1998). They are composed of layers classified in three main groups as input, hidden and output. In each layer there are artificial neuron cells for the aim of processing the data. Because of the layered structure, a particular type of NN is called layered NN. In the layered feed-forward NN, the neurons in a layer are connected to the neurons only in the next layer by adaptive synaptic weights and data flows forward direction. The input neurons receive the input data which are independent variables of the problem. Then the received data is transmitted to the hidden layer neurons by multiplying the corresponding weight values of the connections. The all data entering the hidden neurons are summed by using a chosen summation function for obtaining the net value inside the neuron. After, the net data are activated by an appropriate activation function. The hidden neuron activation function can be theoretically any well-behaved nonlinear function. In this study, tanh (tangent hyperbolic) or ReLU (rectified linear unit) functions have been used for the activations. The advantage of ReLU is its unsaturated gradient, which greatly speeds up the convergence of stochastic gradient landing compared to tanh functions. In the last hidden layer, the data is transmitted to the output layer neurons and NN outputs have been obtained for the dependent variables of the problem. In Fig.1, we have shown the (50-50-50-20) NN structure which is one of the used structures in this study for the determinations of the reaction cross-sections for p-shell stable nuclei. The other used NN structures have been given in Section 3.

The inputs were proton number (Z), neutron number (N) of the target nuclei and photon energy (E) impinging upon the target. Only stable or very-long living isotopes have been considered as target nuclei which are $^7$Li, $^9$Be, $^{10}$Be (1.51x10$^6$ years), $^{10}$B, $^{11}$B, $^{12}$C, $^{13}$C, $^{14}$C (5700 years), $^{14}$N, $^{15}$N and $^{16}$O isotopes. The desired output was photo-neutron reaction cross-section for these different isotopes.

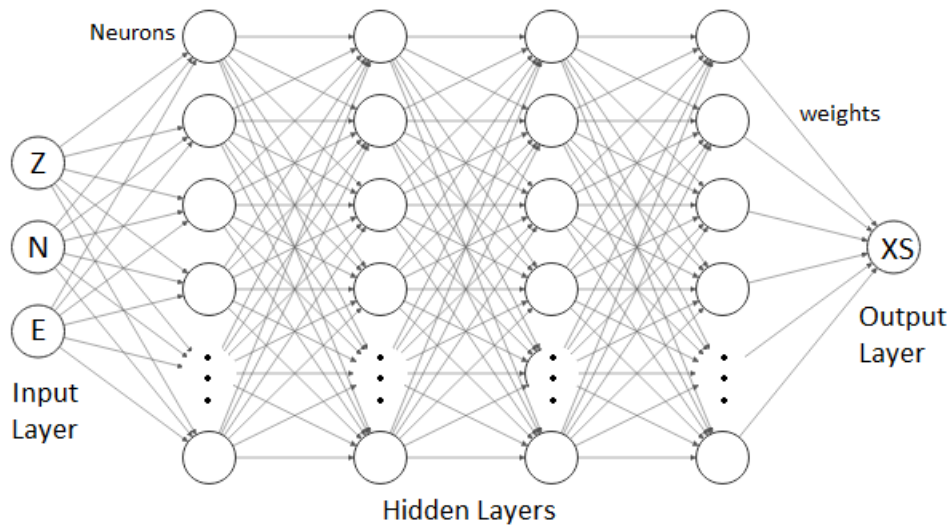

Figure 1. ANN with (50-50-50-20) structure for the prediction of photo-neutron cross sections for p-shell stable target nuclei

The main goal of the method is the determination of the final weight values between neurons by starting random initial values. The NN with best weights gives the NN outputs as close as to the desired values of the problem. In the training stage, NN is trained for the determination of the final best weights by given input and output data values. By the appropriate modifications of the weights, NN modifies its weights until an acceptable error level between NN and desired outputs. The error function was mean square error (MSE) in this study. In the test stage, another dataset of the problem is given to NN and the results are predicted by using the final weights obtained in training process. If the predictions of the test data are well, the ANN is considered to have learned the relationship between input and output data.

In this work, Python programming language for the neural network calculations were used. Python programming language contains fast and practical libraries such as *pandas, numpy, keras*, etc. The data for (γ, n) reaction cross-sections in the literature are studied from threshold energy values to 200 MeV. Total 537 cross-section data has been used for the calculations for p-shell nuclei. All data was divided into three separate sets for training (80%) and test (20%) stages in the calculations. The whole data were obtained from TENDL-2019 reaction cross-section database [6]. The deep sequential neural network model consisting of sequential layers has been used. Each layer added to the deep network is fully connected. In the training stage of NN, the *adam* optimization algorithm (Kingma & Ba, 2017), which is often preferred in deep learning studies, has been used for optimization.

## 3. RESULTS and DISCUSSION

Although there are cross-section values available in the literature, the data do not cover all energy values for target materials. Besides, it is important to have cross-section information for each desired energy values of the photons to be sent on the target materials. Neural network (NN) methods are suitable and easy way for this task. In the calculations of present study, NN method has been employed for the determination of cross-sections whose inputs are atomic number (Z), neutron number (N) of the target material and energy (E) of the incoming photons. Different numbers of hidden layer and neuron have been used which gives the optimal results for their hidden layer configuration classes. These are one hidden layer with 20 neurons, three hidden layers with (3-8-8) configuration, three hidden layers with (50-20-10) configuration, four hidden layers with (50-50-20-20) and five hidden layers with (50-50-20-20-10) configuration, respectively. That is to say, we have got preferable results from 20 neurons for the one hidden layer structure than the other neuron number structures for one hidden layer. For each structure, we have used both ReLU and tanh activation functions separately for the comparison of the results.

After the determination of the final weights in the training, the NN has been first used on the training datasets. According to the results, the best estimation on the training dataset has been obtained for (50-50-50-20) structure with the MSE (minimum square error) value of 0.021 mb. The maximum deviations (MD) from literature data for this NN structure are 0.734 mb for $^{10}$Be at 20 MeV photon energy. In the calculations, ReLU activation function has been used. The corresponding MSE and MD values on the training dataset for tanh activation function are 0.025 mb and 0.867 mb. The MD has been observed for $^{14}$C at 19 MeV photon energy. The MSE value from ReLU activation function are slightly better than the tanh results on the training dataset. The estimations of other NN structures have been shown in Table 1. For ReLU function, the MD have been observed between 1.510 and 9.912 mb for $^{13}$C at 18 MeV, $^{9}$Be at 19 MeV, $^{10}$Be at 19 MeV, $^{9}$Be at 24 MeV and $^{9}$Be at 17 MeV for the NN structure of (20), (3-8-8), (50-20-10), (50-50-20-20) and (50-50-20-20-10). For tanh function, the MD have been observed between 1.336 and 9.248 mb for $^{10}$Be at 20 MeV, $^{11}$B at 18 MeV, $^{14}$C at 17 MeV, $^{14}$C at 15 MeV and $^{10}$Be at 20 MeV for the NN structure of (20), (3-8-8), (50-20-10), (50-50-20-20) and (50-50-20-20-10), respectively.

Table 1. Different structure neural network results for the estimations of cross-sections

| | | *Training* | | *Test* | |
|---|---|---|---|---|---|
| **Hidden neuron number** | **Activation function** | MSE (mb) | MD (mb) | MSE (mb) | MD (mb) |
| 20 | ReLU | 4.473 | 9.771 | 3.555 | 7.227 |
| 3-8-8 | ReLU | 4.767 | 9.912 | 7.563 | 9.984 |
| 50-20-10 | ReLU | 0.840 | 6.107 | 1.099 | 5.925 |
| 50-50-20-20 | ReLU | 0.123 | 2.689 | 2.377 | 7.481 |
| 50-50-50-20 | ReLU | 0.021 | 0.734 | 0.168 | 1.654 |
| 50-50-20-20-10 | ReLU | 0.040 | 1.510 | 1.078 | 7.504 |
| 20 | tanh | 2.688 | 9.248 | 6.005 | 9.973 |
| 3-8-8 | tanh | 3.099 | 9.037 | 3.830 | 9.530 |
| 50-20-10 | tanh | 0.140 | 3.631 | 0.260 | 2.003 |
| 50-50-20-20 | tanh | 0.116 | 2.366 | 0.656 | 6.313 |
| 50-50-50-20 | tanh | 0.025 | 0.867 | 0.258 | 3.271 |
| 50-50-20-20-10 | tanh | 0.024 | 1.336 | 0.325 | 3.174 |

For the seeing of the generalization capability of constructed NN, it has been tested on the test datasets. According to the results, the best predictions on the test dataset have been obtained for the same NN structure with the MSE value of 0.168 mb. The MD from literature data for this NN structure are 1.654 mb for $^{15}$N at 22 MeV photon energy. The corresponding MSE and MD values on the training dataset for tanh activation function are 0.258 mb and 3.271 mb. The MD has been observed for $^{13}$C at 15 MeV photon energy. The MSE value from ReLU activation function are about 1.5 factors better than the tanh results on the test dataset. The predictions of other NN structures have also been shown in Table 1. For ReLU function, the MD have been observed between 5.925 and 9.984 mb for $^{14}$C at 26 MeV, $^{15}$N at 16 MeV, $^{14}$C at 19 MeV, $^{10}$Be at 22 MeV and $^{14}$C at 17 MeV for the NN structure of (20), (3-8-8), (50-20-10), (50-50-20-20) and (50-50-20-20-10). For tanh function, MD have been observed between 2.003 and 9.973 mb for $^{9}$Be at 20 MeV, $^{7}$Li at 22 MeV, $^{14}$N at 16 MeV, $^{14}$C at 18 MeV and $^{9}$Be at 22 MeV for the NN structure of (20), (3-8-8), (50-20-10), (50-50-20-20) and (50-50-20-20-10).

In Figure 2, we have given the best NN predictions of (50-50-50-20) structure with ReLU activation function on the training dataset in comparison with the available literature data. Although the data is highly non-linear, ANN estimations are in harmony with the literature data. The peaks belong to $^{7}$Li, $^{9}$Be, $^{10}$Be, $^{10}$B, $^{11}$B, $^{12}$C, $^{13}$C, $^{14}$C, $^{14}$N, $^{15}$N and $^{16}$O isotopes. The largest cross-section has been obtained for $^{14}$C isotopes with its maximum value of 33.5 mb at 17 MeV energy value. Its literature value is 33.1 mb. The smallest cross-section has been seen for $^{12}$C isotopes. The maximum of the cross-section for this isotope is 2.03 mb at 22 MeV whereas the literature value is 2.00 mb.

The maximum cross-section values are 10.23 mb at 22 MeV for $^{7}$Li, 14.66 mb at 20 MeV for $^{9}$Be, 26.70 mb at 19 MeV for $^{10}$Be, 9.00 mb at 19 MeV for $^{10}$B, 12.33 mb at 18 MeV for $^{11}$B, 2.03 mb at 22 MeV for $^{12}$C, 17.17 mb at 18 MeV for $^{13}$C, 33.45 mb at 17 MeV for $^{14}$C, 3.28 mb at 17 MeV for $^{14}$N, 16.96 mb at 17 MeV for $^{15}$N and 0.96 mb at 17 MeV for $^{16}$O. Whereas the literature values are 10.72, 14.99, 27.33, 9.05, 12.47, 2.00, 16.95, 33.10, 2.96, 16.93 and 0.96, respectively. The cross-sections get their maximums for the nuclei between 17-22 MeV in the investigated energy range from threshold energies to 200 MeV. The reaction thresholds are 8, 2, 7, 9, 12, 19, 5, 9, 11, 11 and 16 MeV for $^{7}$Li, $^{9}$Be, $^{10}$Be, $^{10}$B, $^{11}$B, $^{12}$C, $^{13}$C, $^{14}$C, $^{14}$N, $^{15}$N and $^{16}$O isotopes, respectively.

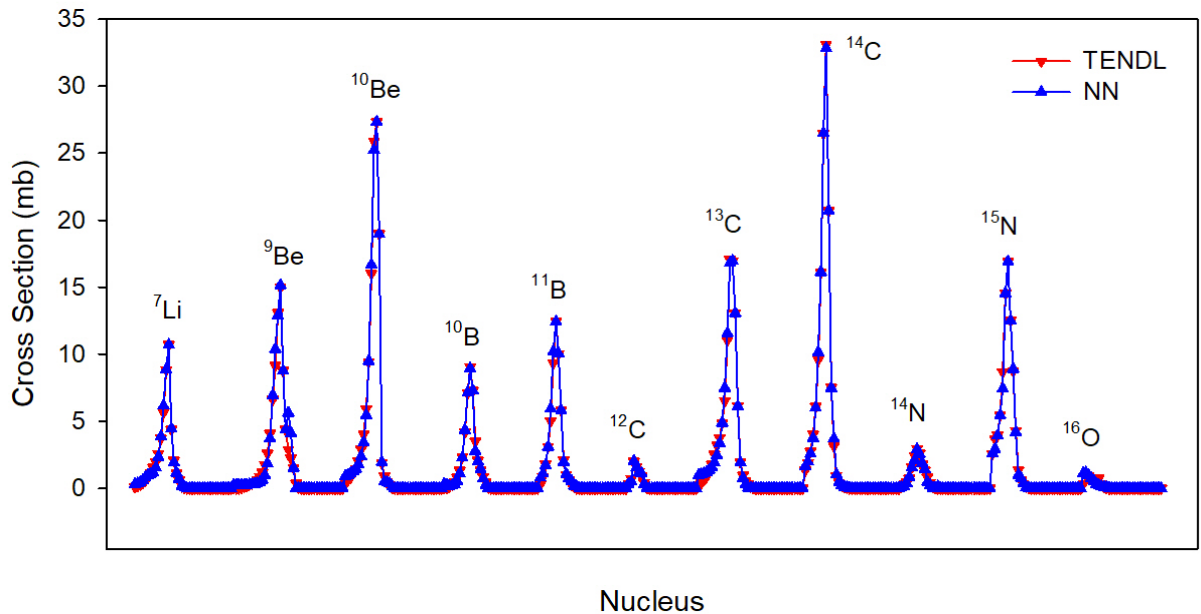

Figure 2. Literature (TENDL) data and best NN estimations with (50-50-50-20) structure on photo-neutron reaction cross-section on stable p-shell nuclei (top) and differences between them (bottom)

In Figures 3-7, we have given the differences between the NN predictions and the literature values on relevant cross-section data. These have been shown for both training and test datasets separately for either ReLU or tanh activation functions.

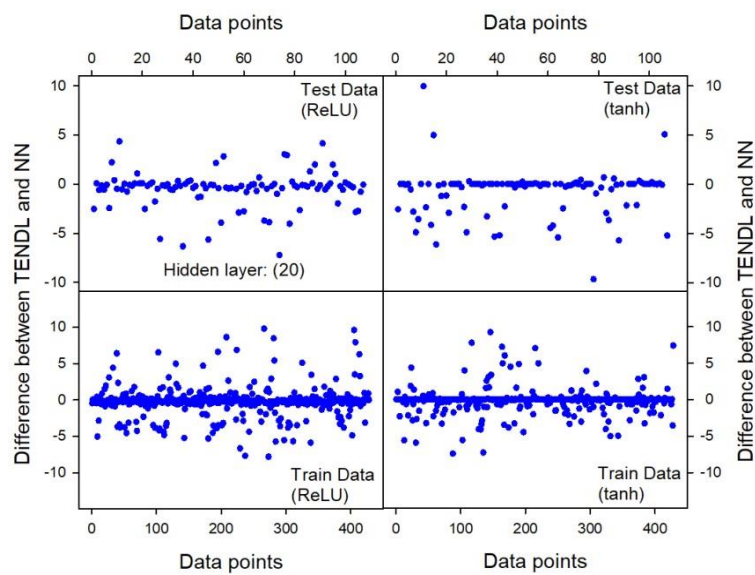

Figure 3. Difference between literature (TENDL) data and NN (20) estimations on test (top) and train (bottom) datasets with ReLU (left) and tanh (right) functions

For the 20 neurons in one hidden layer NN structure, the estimations on the training data for tanh activation function are better than the ReLU results. Namely, the training of the NN has been performed better for tanh, whereas the test of the NN is slightly worst (Figure 3). However, it is not appropriate to use this NN structure since the estimates are spread around 10 mb. For the (3-8-8) hidden layer configuration of NN, tanh activation function gives slightly better results on both train and test datasets (Figure 4). But since the estimates still reach around 10 mb, this structure is also not suitable for use.

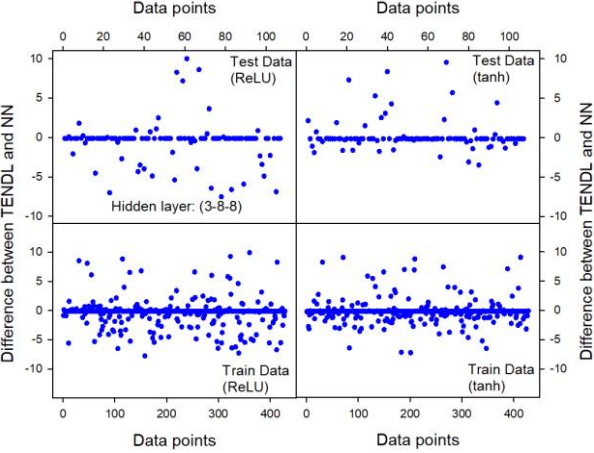

Figure 4. Difference between literature (TENDL) data and NN (3-8-8) estimations on test (top) and train (bottom) datasets with ReLU (left) and tanh (right) functions

For the (50-20-10) hidden layer configuration of NN which is larger in terms of neuron numbers, the estimations for tanh activation function are better than the ReLU results. The results are 6 and 4 factors better for train and test datasets, respectively (Figure 5). The deviations for predictions on test datasets are between -2 and 2 mb indicate that the larger structures become convenient for the problem. For the (50-50-20-20) hidden layer configuration of NN, the estimations for tanh activation function are slightly better than the ReLU results on the train dataset. Furthermore, the predictions on the test datasets with tanh function are 3.6 factors better (Figure 6). Still, the NN structure should be improved for the good estimations on the cross-section data especially for ReLU.

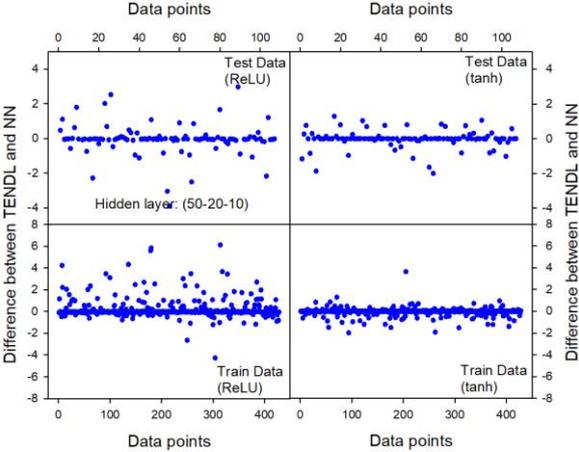

Figure 5. Difference between literature (TENDL) data and NN (50-20-10) estimations on test (top) and train (bottom) datasets with ReLU (left) and tanh (right) functions

For the (50-50-50-20) hidden layer configuration of NN, the estimations for ReLU activation function are somewhat better than the tanh results on both train and test datasets. The results are 6 and 4 factors better for train and test datasets, respectively (Figure 7). It is clear in the figure that the predictions are concentrated between -1 and 1 mb. The best results have been obtained by using this NN structure.

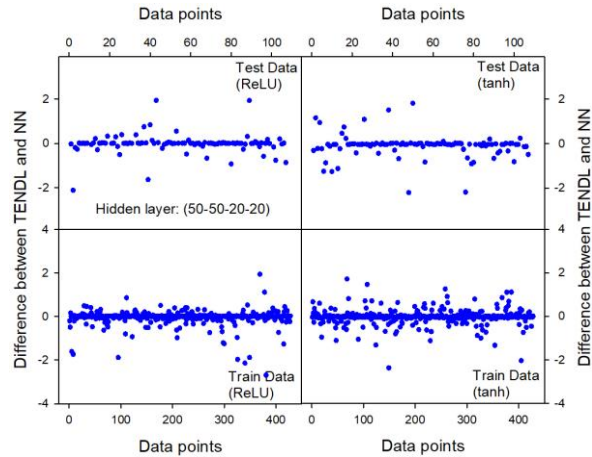

Figure 6. Difference between literature (TENDL) data and NN (50-50-20-20) estimations on test (top) and train (bottom) datasets with ReLU (left) and tanh (right) functions

Lastly, we have tried to larger hidden layer number structure with the (50-50-20-20-10) configuration. For this NN, the training has been performed better by using ReLU activation function than tanh. The estimations on train dataset are 6 factors better than the estimations by using tanh. Whereas for the predictions on test dataset, tanh gives 3.3 factors better results than those of ReLU (Figure 8). Using more than four hidden layers causes results to get worse again.

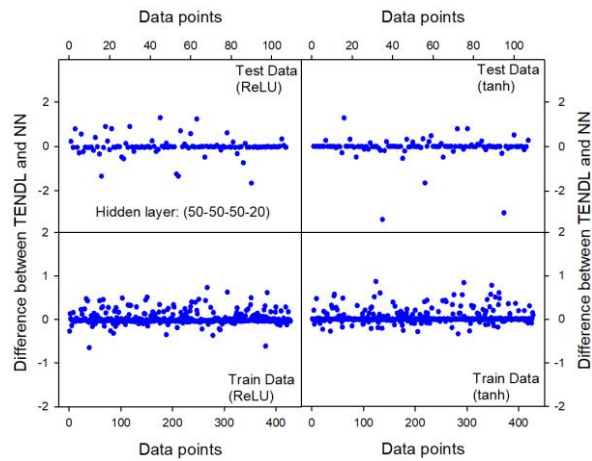

Figure 7. Difference between literature (TENDL) data and NN (50-50-50-20) estimations on test (top) and train (bottom) datasets with ReLU (left) and tanh (right) functions

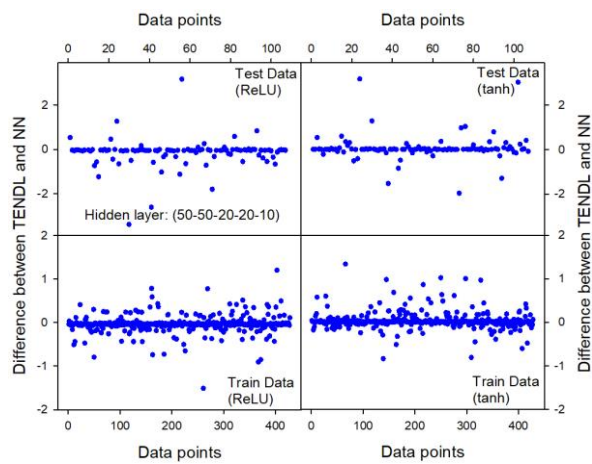

Figure 8. Difference between literature (TENDL) data and NN (50-50-20-20-10) estimations on test (top) and train (bottom) datasets with ReLU (left) and tanh (right) functions

## 4. CONCLUSIONS

In this work, (γ, n) photo-neutron reaction cross-sections for the stable or long-lived isotopes in p-shell have been predicted by using neural network (NN) methods with the different hidden layer and neuron combinations in the threshold to 200 MeV energy range. The results have been compared with each other and the available literature data. The data for the applications of the methods have been borrowed from TENDL-2019 nuclear data library. According to the results, the predictions for the cross-sections are very close to the available literature data. Therefore, one can use the NN methods for the obtaining of photo-neutron reaction cross-sections whose values are not available in the literature. In detail, the increase in the number of hidden layers used and the number of hidden neurons generally improves the results. The obtained better results have generally been come from the activation function of tanh. But the present problem, (50-50-50-20) hidden layer configuration in four hidden layer with ReLU function have given the best results. The use of four hidden layers (deep neural network) with many neurons is more suitable for the obtaining of photo-neutron reaction cross-sections on p-shell nuclei.

## References

Akkoyun, S. (2013). Time-of-flight discrimination between gamma-rays and neutrons by using artificial neural networks. *Annals of Nuclear Energy*, *55*, 297-301. https://doi.org/10.1016/j.anucene.2013.01.006

Akkoyun, S., Bayram, T., Kara, S. O., & Sinan, A. (2013). An artificial neural network application on nuclear charge radii. *Journal of Physics G: Nuclear and Particle Physics*, *40*(5), 055106. https://doi.org/10.1088/0954-3899/40/5/055106

Akkoyun, Serkan. (2020). Estimation of fusion reaction cross-sections by artificial neural networks. *Nuclear Instruments and Methods in Physics Research Section B: Beam Interactions with Materials and Atoms*, *462*, 51-54. https://doi.org/10.1016/j.nimb.2019.11.014

Akkoyun, Serkan, Bayram, T., Dulger, F., Đapo, H., & Boztosun, I. (2016). Energy level and half-life determinations from photonuclear reaction on Ga target. *International Journal of Modern Physics E*, *25*(08), 1650045. https://doi.org/10.1142/S0218301316500452

Akkoyun, Serkan, Bayram, T., & Turker, T. (2014). Estimations of beta-decay energies through the nuclidic chart by using neural network. *Radiation Physics and Chemistry*, *96*, 186-189. https://doi.org/10.1016/j.radphyschem.2013.10.002

Akkoyun, Serkan, Bayram, T., & Yildiz, N. (2016). Estimations of Radiation Yields for Electrons in Various Absorbing Materials. *Cumhuriyet Üniversitesi Fen-Edebiyat Fakültesi Fen Bilimleri Dergisi*, *37*, 59-65.

Akkoyun, Serkan, & Kara, S. O. (2013). An approximation to the cross sections of Zlboson production at CLIC by using neural networks. *Central European Journal of Physics*, *11*(3), 345-349. https://doi.org/10.2478/s11534-012-0168-y

Bayram, T., Akkoyun, S., & Şentürk, Ş. (2018). Adjustment of Non-linear Interaction Parameters for Relativistic Mean Field Approach by Using Artificial Neural

Networks. *Physics of Atomic Nuclei*, *81*(3), 288-295. https://doi.org/10.1134/S1063778818030043

Bayram, Tuncay, Akkoyun, S., & Kara, S. O. (2014). A study on ground-state energies of nuclei by using neural networks. *Annals of Nuclear Energy*, *63*, 172-175. https://doi.org/10.1016/j.anucene.2013.07.039

Chadwick, J., & Goldhaber, M. (1934). A Nuclear Photo-effect: Disintegration of the Diplon by -Rays. *Nature*, *134*(3381), 237-238. https://doi.org/10.1038/134237a0

Haykin, S. (1998). *Neural Networks: A Comprehensive Foundation* (2 edition). Prentice Hall.

Ishkhanov, B. S., & Orlin, V. N. (2009). Description of cross sections for photonuclear reactions in the energy range between 7 and 140 MeV. *Physics of Atomic Nuclei*, *72*(3), 410-424. https://doi.org/10.1134/S1063778809030041

Kara, S. O., Akkoyun, S., & Bayram, T. (2014). Probing for leptophilic gauge boson Zl at ILC with $\sqrt{s} = 1~{\rm TeV}$ by using ANN. *International Journal of Modern Physics A*, *29*(30), 1450171. https://doi.org/10.1142/S0217751X14501711

Kingma, D. P., & Ba, J. (2017). Adam: A Method for Stochastic Optimization. *arXiv:1412.6980 [cs]*. http://arxiv.org/abs/1412.6980

Koning, A. J., Rochman, D., Sublet, J.-Ch., Dzysiuk, N., Fleming, M., & van der Marck, S. (2019). TENDL: Complete Nuclear Data Library for Innovative Nuclear Science and Technology. *Nuclear Data Sheets*, *155*, 1-55. https://doi.org/10.1016/j.nds.2019.01.002

Strauch, K. (1953). Recent Studies of Photonuclear Reactions. *Annual Review of Nuclear Science*, *2*(1), 105-128. https://doi.org/10.1146/annurev.ns.02.120153.000541

Utsuno, Y., Shimizu, N., Otsuka, T., Ebata, S., & Honma, M. (2015). Photonuclear reactions of calcium isotopes calculated with the nuclear shell model. *Progress in Nuclear Energy*, *82*, 102-106. https://doi.org/10.1016/j.pnucene.2014.07.036

Yildiz, N., & Akkoyun, S. (2013). Neural network consistent empirical physical formula construction for neutron–gamma discrimination in gamma ray tracking. *Annals of Nuclear Energy*, *51*, 10-17. https://doi.org/10.1016/j.anucene.2012.07.042

Yildiz, N., Akkoyun, S., & Kaya, H. (2018). Consistent Empirical Physical Formula Construction for Gamma Ray Angular Distribution Coefficients by Layered Feedforward Neural Network. *Cumhuriyet Science Journal*, *39*(4), 928-933. https://doi.org/10.17776/csj.476733